\documentclass[twocolumn,showpacs,aps,prl,letter,amsmath,amssymb,superscriptaddress]{revtex4-1}
\usepackage{bm,color,bbm}
\usepackage{hyperref, mathtools,graphicx,natbib}

\newcommand{\beq}{\begin{equation}}
\newcommand{\eeq}{\end{equation}}
\newcommand{\bqa}{\begin{eqnarray}}
\newcommand{\eqa}{\end{eqnarray}}
\newcommand{\nn}{\nonumber}

\newcommand{\rt}[1]{\sqrt{#1}\,}

\newcommand{\smallfrac}[2]{\mbox{$\frac{#1}{#2}$}}
\newcommand{\bra}[1]{ \langle{#1} |}
\newcommand{\ket}[1]{ |{#1} \rangle}

\newcommand{\half}{\smallfrac{1}{2}}

\newcommand{\sq}[1]{\left[ {#1} \right]}

\newcommand{\tr}[1]{{\rm Tr}\sq{ {#1} }}

\newcommand{\mf}{\mathbf}

\definecolor{maroon}{rgb}{0.7,0,0}

\definecolor{ngreen}{rgb}{0.3,0.7,0.3}

\definecolor{golden}{rgb}{0.8,0.6,0.1}

\begin{document}
\title{Anisotropic invariance and the distribution of quantum correlations}
\author{Shuming Cheng}
\affiliation{Centre for Quantum Computation and Communication Technology (Australian Research Council), Centre for Quantum Dynamics, Griffith University, Brisbane, QLD 4111, Australia}
\affiliation{Key Laboratory of Systems and Control, Academy of Mathematics and Systems Science, Chinese Academy of Sciences, Beijing 100190, P. R. China}
\author{Michael J. W. Hall}
\affiliation{Centre for Quantum Computation and Communication Technology (Australian Research Council), Centre for Quantum Dynamics, Griffith University, Brisbane, QLD 4111, Australia}
\begin{abstract}

We report the discovery of two new invariants for three-qubit states which, similarly to the 3-tangle, are invariant under local unitary transformations and permutations of the parties.  These quantities have a direct interpretation in terms of the anisotropy of pairwise spin correlations. Applications include a universal ordering of pairwise quantum correlation measures for pure three-qubit states; tradeoff relations  for anisotropy, 3-tangle  and Bell nonlocality; strong monogamy relations for Bell inequalities, Einstein-Podolsky-Rosen steering inequalities, geometric discord and fidelity of remote state preparation  (including results for arbitrary three-party states); and a statistical and reference-frame-independent form of quantum secret sharing.  
\end{abstract}

\maketitle

Quantum systems of just two qubits are sufficient for exploring a remarkable wealth of quantum phenomena~\cite{NC00}, and yet cannot be said to be fully understood.  Studying systems of three or more qubits is, accordingly, expected to yield even richer returns.  Indeed, phenomena such as quantum teleportation \cite{Teleport93}, genuine multipartite entanglement \cite{HHHH09}, quantum monogamy \cite{CKW00}, decoherence \cite{Zeh70, Z81}, Wigner's friend \cite{W61} and the Greenberger-Horne-Zeilinger (GHZ) paradox \cite{GHZ90, M90} can only arise for tripartite and higher-order systems. 

 In this paper we note that the spin correlations of any two-qubit state decompose naturally into isotropic and anisotropic components, and introduce corresponding measures of isotropy and anisotropy that are invariant under local unitary and antiunitary transformations.  Surprisingly, for pure three-qubit states the anisotropy properties are found to be identical for each pair, i.e., they have the same values when evaluated for any two of the qubits.  This pairwise invariance provides a crucial link, hitherto missing, that connects fundamental properties of two-party and three-party quantum correlations. 
  
We give a number of applications of anisotropic invariance, the first being  a simple and universal answer to an entanglement-ordering question: Which of the three reduced pairs of a pure three-qubit state is more strongly entangled, more strongly EPR-steerable, or more strongly Bell nonseparable? The answer further leads to the elegant confirmation of a recently conjectured complementarity relation for bipartite Bell nonlocality and 3-tangle \cite{PMC15}, and also to a tradeoff relation between Bell nonlocality and anisotropy. 

 We then derive  strong quantum monogamy relations for Bell nonlocality \cite{BCPSW14} and Einstein-Podolsky-Rosen (EPR) steering \cite{WJD07}, beyond those in the literature.
For example, consider an arbitrary three-qubit state shared by three parties Alice, Bob, and Charlie in spacelike separated regions. If Alice and Bob choose to measure spin in directions $\bm a_1$ or $\bm a_2$ and $\bm b_1$ or $\bm b_2$, respectively, they can determine the average value of the Bell operator  $\mathcal B_{AB}(\bm a_1,\bm a_2,\bm b_1,\bm b_2)=\bm \sigma\cdot \bm a_1\otimes \bm \sigma \cdot \bm b_1+\bm \sigma\cdot \bm a_1\otimes \bm \sigma\cdot \bm b_2+\bm \sigma\cdot \bm a_2\otimes \bm \sigma\cdot \bm b_1-\bm \sigma\cdot \bm a_2\otimes \bm \sigma\cdot \bm b_2$. A violation of the well known Clauser-Horne-Shimony-Holt (CHSH)  inequality, $\langle\mathcal B_{AB}(\bm a_1,\bm a_2,\bm b_1,\bm b_2)\rangle^2\leq 4$, implies that Alice and Bob share  correlations that cannot be explained by any local hidden variable model \cite{CHSH69,BCPSW14}.  We show that 
\begin{equation} \label{chshmon}
\langle\mathcal B_{AB}(\bm a_1,\bm a_2,\bm b_1,\bm b_2)\rangle^2 + \langle\mathcal B_{AC}(\bm a'_1,\bm a'_2,\bm c_1,\bm c_2)\rangle^2 \leq 8 ,
\end{equation}
for any choices of the measurement directions. Thus, at most one pair can violate the CHSH inequality---even 
if each pair is allowed to optimise over all possible measurement directions. This is stronger than previous monogamy relations, which either give weaker restrictions on the expectation values \cite{SG01,SGPRA01}; or require fixed measurement directions $\bm a_1=\bm a'_1$, $\bm a_2=\bm a'_2$ for Alice \cite{TV06, BCPSW14}. The strength of Eq.~(\ref{chshmon}) is unexpected---indeed it has recently been claimed, incorrectly, that this monogamy relation can be explicitly violated \cite{QFL15}. 

We obtain similar three-qubit monogamy relations for witnessing EPR steering \cite{CJWR09, ECMH15}; for geometric discord \cite{DVB10}; and for the fidelity of remote state preparation~\cite{pati,Bennett,DB12}. Further, Eq.~(\ref{chshmon}) is used to substantially simplify the derivation of a general monogamy relation for Bell nonlocality, for arbitrary three-party quantum systems \cite{TV06}. We also derive a strong monogamy relation for the EPR steering of a qubit by arbitrary quantum systems.

 Finally,  we show that anisotropic invariance provides a  noise-robust  statistical form of quantum secret sharing \cite{HBB99}, in which any two of three parties can cooperate to recover information coded into an ensemble of three-qubit states, without any need to establish a common reference frame;  and note the relevance of anisotropic invariance to the quantum marginal problem~\cite{K04}.

\paragraph{Isotropic vs anisotropic spin correlations.---}
Any two-qubit state $\rho_{AB}$ can be written in  the form 
	\begin{align}
		\rho_{AB} =  \frac{1}{4} \bigg( & \mathbbm{1}_A \otimes \mathbbm{1}_B+{\mf{a}} \cdot \boldsymbol{\sigma} \otimes \mathbbm{1}_B+\mathbbm{1}_A \otimes {\mf{b}} \cdot \boldsymbol{\sigma}     \nn \\  
		&  \qquad\qquad+\sum^3_{j,k=1}T_{jk}\,\sigma_j\otimes\sigma_k \bigg), 	\label{state}
	\end{align}
	where $\boldsymbol{\sigma}\equiv (\sigma_1, \sigma_2, \sigma_3)$ denotes the vector of Pauli spin operators, $\mathbbm{1}_A,\mathbbm{1}_B$ are identity operators,  ${\mf{a}}$ and ${\mf{b}}$ are the Bloch vectors of Alice's and Bob's reduced states, and $T$ is the spin correlation matrix, i.e., $T_{jk}=\langle\sigma_j\otimes \sigma_k\rangle$.
	
	The spin matrix $T$ plays a fundamental role in encoding the global information of the two-qubit system, and is deeply connected to the strength of its quantum correlations~\cite{HHHH09,MBCPV12,BCPSW14}. 
In particular, the eigenvalues of the symmetric matrix ${\cal S}=TT^\top$ (or, equivalently, of $T^\top T$), are invariant under  arbitrary rotations and reflections of the parties' local coordinate systems, as well as under interchange of the qubits. These eigenvalues determine, for example, the maximal possible violation of the CHSH inequality \cite{HHH95} and of various EPR steering inequalities \cite{GC16, CA16,QZFY16}, as well as the maximum fidelity of remote state preparation \cite{DB12} (see also below). 

We will denote the eigenvalues of ${\cal S}$ by $s_1\geq s_2\geq s_3\geq 0$. They are all equal if and only if the spin correlations are isotropic, relative to suitable local coordinates, such as for a two-qubit Werner state \cite{W89}. A natural measure of the degree of  isotropy is the average quantity 
\begin{align} 
s_{\rm iso}& := \frac{1}{3}\sum_j s_j, \label{siso}
\end{align} 
which we will refer to as the isotropic strength.  For a Werner state, comprising a fraction $W$ of the singlet state and fraction $1-W$ of the maximally mixed state, the isotropic strength is easily determined to be $s_{\rm iso}=W^2$.  

It follows that the differences $\delta s_j:=s_j - s_{\rm iso}$ quantify the  inherent  {\it anisotropy} of the spin correlations. As these differences sum to zero, there are two independent measures of spin anisotropy---which we may take, for example, to be the eigenvalue gaps
\begin{align} \label{gaps}
g_1 :=  s_1-s_2 ,
\qquad g_2:=s_2-s_3 . 
\end{align} 
Alternative, more symmetric  choices are the anisotropic strength and the anisotropic volume, defined by
\begin{align} \nn
s_{\rm ani}^{~2} := \sum_j (\delta s_j)^2 &= \frac{(s_1-s_2)^2 + (s_2-s_3)^2+(s_3-s_1)^2}{3},\\
V_{\rm ani} &:= \delta s_1\,\delta s_2\,\delta s_3, \label{vani}
\end{align}
respectively.
Note that $V_{\rm ani}$  is a signed volume measure. 
These  measures also arise naturally via the unique decomposition of the matrix  ${\cal S}=TT^\top$ into isotropic and anisotropic components, ${\cal S}=s_{\rm iso}I + {\cal S}_{\rm ani}$, with $s_{\rm ani}=({\rm tr}[{\cal S}_{\rm ani}^{~2}])^{1/2}$ and $V_{\rm ani}=\det {\cal S}_{\rm ani}$, as shown in the Supplemental Material \cite{SM}.
We emphasize that all measures in Eqs.~(\ref{siso})-(\ref{vani}) are invariant under local rotations and reflections, and hence they can be experimentally determined whether or not the two parties share a common reference frame.  The advantage of distinguishing between isotropic and anisotropic contributions becomes apparent as soon as more qubits are considered.

\paragraph{Isotropic and anisotropic invariants for three qubits.---}
For a pure three-qubit state $\rho_{ABC}=\ket{\Psi_{ABC}}\bra{\Psi_{ABC}}$ shared by Alice, Bob, and Charlie, the pairwise correlations are described by the reduced density operators $\rho_{AB}$, $\rho_{AC}$, and $\rho_{BC}$,  each of which can be decomposed similarly to Eq.~(\ref{state}). The Bloch vectors for Alice, Bob, and Charlie will be denoted by $\mf{a}, \mf{b},$ and $\mf{c}$, respectively, and the corresponding pairwise spin correlation matrices by $T^{AB}$, $T^{AC},$ and $T^{BC}$. 

The isotropic strengths of the three two-qubit states may be calculated from Eq.~(\ref{siso}), and need not be equal.  However, their sum is constrained to be constant:
\beq
s_{\rm iso}^{AB}+s_{\rm iso}^{AC}+s_{\rm iso}^{BC}=1, \label{isotropy}
\eeq
as follows immediately from Eq.~(\ref{siso}) and the property ${\rm tr}[\mathcal{S}^{AB}+\mathcal{S}^{AC}+\mathcal{S}^{BC}]=3$ \cite{QFL15}. This property, and hence Eq.~(\ref{isotropy}), is easily established by equating the purities of $\rho_A$ and $\rho_{BC}$, and similarly for other bipartitions, and generalises to an inequality for mixed three-qubit states~\cite{CMHW16}. 

Remarkably, in strong contrast to the isotropic strengths, we have found that the spin anisotropy properties are identical for each pair of qubits: 
\begin{align}
\delta s^{AB}_j=\delta s^{AC}_j=\delta s^{BC}_j,\qquad j=1,2,3. \label{aniso1} 
\end{align} 
 It follows that the anisotropy measures in Eqs.~(\ref{gaps})-(\ref{vani}) are similarly identical. This highly nontrivial property is proved in the Supplemental Material,  and  can be generalised  to mixed states via a convex-roof construction~\cite{SM}.

Note from Eq.~(\ref{aniso1}) that the pairwise anisotropies are not only invariant under local transformations, but under any permutation of the qubits or parties.  This is reminiscent of the 3-tangle, $\tau$, which for pure three-qubit states can also be calculated from any one of the pairwise reduced states, and  is similarly invariant both under local transformations and permutations \cite{CKW00, BL16}. Importantly, however, spin anisotropy  is independent of the 3-tangle.  For example, there are classes of states for which $\tau$ varies over all possible values, but for which $\delta s_j$ and the anisotropy measures in Eqs.~(\ref{gaps}) and~(\ref{vani}) are constant, and  vice versa  \cite{SM}.  Noting that pure three-qubit states are characterised, up to local unitary transformations, by six invariants \cite{AACJLT00,AAER01}, it follows that a particularly symmetric choice for these is given by supplementing the Bloch vector lengths $a$, $b$, $c$ (or the concurrences) by the 3-tangle $\tau$ and the anisotropic strength $s_{\rm ani}$ and volume $V_{\rm ani}$. 

\paragraph{Entanglement ordering.---}
As a first application, we consider the question posed earlier: Which of the three pairs of a three-qubit pure state is more strongly entangled, more strongly EPR-steerable, or more strongly Bell nonlocal? We show that the same answer can be given in each case, corresponding to the pair which has the greatest spin isotropy.

In particular, for the bipartite state $\rho_{AB}$ shared by Alice and Bob, consider the entanglement as measured by the concurrence ${\mathcal C}^{AB}$ \cite{HW97, W98}, and the degree of Bell nonlocality as measured by the Horodecki parameter~\cite{HHH95},
\beq \label{mab}
{\mathcal M}^{AB}=s_1^{AB}+s_2^{AB} .
\eeq  
Note ${\mathcal M}^{AB}$ also quantifies the degree of EPR-steerability witnessed via two spin measurements by each party   \cite{GC16, CA16,QZFY16}. Then, as shown in the Supplemental Material, the triples
$({\mathcal C}^{AB}, {\mathcal C}^{AC}, {\mathcal C}^{BC})$ and $({\mathcal M}^{AB}, {\mathcal M}^{AC}, {\mathcal M}^{BC})$
each have the same ordering as $(s_{\rm iso}^{AB},s_{\rm iso}^{AC}, s_{\rm iso}^{BC})$, for any pure three-qubit state \cite{SM}. For example, 
\beq 
{\mathcal C}^{AB}\geq {\mathcal C}^{AC}~~{\rm iff~~} {\mathcal M}^{AB}\geq {\mathcal M}^{AC}~~{\rm iff~~} s_{\rm iso}^{AB}\geq s_{\rm iso}^{AC} . \label{entangle ordering}
\eeq 
 It is also shown  that the triple $(c,b,a)$ of Bloch vector lengths (and hence the purities of $\rho_{AB}$, $\rho_{AC}$ and $\rho_{BC}$) has the same ordering~\cite{SM}.

The above ordering of pairwise correlation strengths arises from the stronger quantitative relations
\begin{align} \nn
	\left({\mathcal C}^{AB}\right)^2 - \left({\mathcal C}^{AC}\right)^2 
	&= \half({ {\mathcal M}^{AB}-{\mathcal M}^{AC}}) 
	=  s_{\rm iso}^{AB}-s_{\rm iso}^{AC} \\
	&=c^2-b^2 \label{ab},
\end{align}
and its permutations.  These  are simple consequences of Eqs.~(\ref{siso}) and~(\ref{aniso1}), as shown in the Supplemental Material~\cite{SM}.  The relation between ${\mathcal M}^{AB}-{\mathcal M}^{AC}$ and $c^2-b^2$ extends Theorem~2  in Ref.~\cite{SDSS16} to all pure three-qubit states. Note from  Eq.~(\ref{ab})  that the relative entanglement ordering can be determined by local measurements of  the Bloch vector  lengths.

\paragraph{Tradeoffs for Bell nonlocality.---}

Equation~(\ref{ab}) further leads to the simple verification of a recently conjectured complementarity relation~\cite{PMC15}
\beq
\max \{{\mathcal M}^{AB}, {\mathcal M}^{AC}, {\mathcal M}^{BC}\}+\tau \leq 2, \label{complementarity}
\eeq
for general three-qubit states (see Supplemental Material~\cite{SM}).  Recalling that ${\mathcal M}^{AB}>1$ is necessary for Alice and Bob to be able to violate the CHSH inequality  \cite{HHH95}, this imposes a strong tradeoff between Bell nonlocality and 3-tangle. 

A similar tradeoff relation can also be obtained for Bell nonlocality and anisotropy:
\beq \label{trade}
\max \{{\mathcal M}^{AB}, {\mathcal M}^{AC}, {\mathcal M}^{BC}\}+g_1 + g_2 \leq 2  .
\eeq
This holds for all pure three-qubit states, and may be generalised to mixed states via a convex-roof extension of the eigenvalue gaps in Eq.~(\ref{gaps}) (see Supplemental Material~\cite{SM}).  For example, a pure state with maximum anisotropy $g_1+g_2=\delta s_1-\delta s_3=1$, such as the GHZ state $\frac{1}{\sqrt{2}}(|000\rangle+|111\rangle)$, cannot violate a CHSH inequality.

\paragraph{Strong three-qubit monogamy relations.---}

The invariance of spin anisotropy as per Eq.~(\ref{aniso1}) allows us to rewrite the Horodecki parameter in Eq.~(\ref{mab}) as 
\begin{align} 
{\mathcal M}^{AB}	&= (s_{\rm iso}^{AB}+\delta s^{AB}_1)+(s_{\rm iso}^{AB}+\delta s^{AB}_2) \nn \\
&= 2s_{\rm iso}^{AB}-\delta s^{AB}_3 \nn \\
&= 2s_{\rm iso}^{AB}+(\delta s^{AB}_3- \delta s^{AC}_3-\delta s^{BC}_3)\nn \\
&= s_{\rm iso}^{AB}+ s_{\rm iso}^{AC}+s_{\rm iso}^{BC}+(s_{\rm iso}^{AB}+ \delta s^{AB}_3) \nn \\
&\qquad-(s_{\rm iso}^{AC}+ \delta s^{AC}_3) -(s_{\rm iso}^{BC}+ \delta s^{BC}_3)  \nn\\
&= 1+s_3^{AB}-s_3^{AC}-s^{BC}_3 \label{nice}
\end{align}
for pure three-qubit states, where the last line follows via Eq.~(\ref{isotropy}). 
Similar expressions for ${\mathcal M}^{AC}$ and ${\mathcal M}^{BC}$ are obtained by permuting the labels, yielding
\beq \label{monog}
{\mathcal M}^{AB}+{\mathcal M}^{AC}=2(1-s_3^{BC})\leq 2.
\eeq
The strong monogamy relation in Eq.~(\ref{chshmon}) immediately follows for such states, noting that $4{\mathcal M}^{AB}$ is the maximum possible expectation value of the CHSH parameter $\langle\mathcal B_{AB}(\bm a_1,\bm a_2,\bm b_1,\bm b_2)\rangle^2$  \cite{HHH95}.  
The generalisation of Eqs.~(\ref{chshmon}) and~(\ref{monog}) to mixed three-qubit states $\rho_{ABC}=\sum_n p_n|\psi_n\rangle\langle \psi_n|$ is a trivial consequence of the convexity property
 \begin{align} \nn
 \langle\mathcal B_{AB}\rangle^2 \!=\! \left(\sum_n p_n\langle\psi_n|\mathcal B_{AB}|\psi_n\rangle\right)^2 \!\leq \sum_n p_n \langle\psi_n|\mathcal B_{AB}|\psi_n\rangle^2.
 \end{align}

The condition ${\mathcal M}^{AB}> 1$ is  sufficient  not only for Alice and Bob to be able to witness violation of the CHSH inequality, but also for them to be able to  violate various EPR steering inequalities via two spin measurements each  \cite{GC16, CA16,QZFY16}. 
Hence, inequality~(\ref{monog}) also imposes strict monogamy on quantum steering: at most one pair can  violate these inequalities,  even when each pair is free to optimise their choice of measurement directions. This improves on the steering monogamy relation in Eq.~(18) of Reid ~\cite{R13} (for $m=2)$, which requires two fixed measurement directions for each party.

We can also obtain monogamy relations for weaker forms of quantum correlation. For example, the geometric quantum discord between two qubits is given by $\mathcal{D}_{A\rightarrow B}=\frac{1}{4}\Big(a^2+{\rm tr}[T^{AB}(T^{AB})^\top]-k_{\max}\Big) \leq\half$, 
where $k_{\max}=\max_{|\mathbf{x}|\leq1}(\mathbf{x}\cdot\mathbf{a})^2+\mathbf{x}^\top T^{AB}(T^{AB})^\top\mathbf{x}$~\cite{DVB10}. Making the nonoptimal choice $\mathbf{x}=\mathbf{a}$ yields the upper bound  $\mathcal{D}_{A\rightarrow B}\leq\frac{1}{4}(s_1^{AB}+s^{AB}_2)=\frac{1}{4} {\mathcal M}^{AB},$ 
leading via Eq.~(\ref{monog}) to the monogamy relation
 \beq
 \mathcal{D}_{A\rightarrow B}+\mathcal{D}_{A\rightarrow C} \leq \half \label{discord}
 \eeq
 for the nonclassicality of general 3-qubit states,  as measured by the geometric discord. A  stronger  relation,  but valid  for pure states only, is given in Ref.~\cite{SAPB12}. 

Moreover, for a two-qubit state $\rho_{AB}$, the remote state preparation fidelity is given by $\mathcal{F}^{AB}=\frac{1}{2}\left(s^{AB}_2+s^{AB}_3\right)$ \cite{DB12}. Thus, $\mathcal{F}^{AB}\leq \half {\mathcal M}^{AB}$, immediately yielding the corresponding monogamy relation 
\beq
\mathcal{F}^{AB}+\mathcal{F}^{AC}\leq 1.
\eeq
Hence, the ability of Alice to remotely prepare the state of Bob's qubit is strongly constrained by her ability to prepare the state of Charlie's qubit.  We note it is easy to show,  for  the case of pure states, that the fidelities have the same ordering as in Eq.~(\ref{entangle ordering})~\cite{SM}.

\paragraph{Strong general monogamy relations.---} If Alice, Bob and Charlie share a general tripartite quantum system, rather than three qubits, then it may be possible for each pair to make local measurements with outcomes $\pm1$ that violate the CHSH inequality. A simple example is the six-qubit state $|\Psi\rangle_{AB}\otimes|\Psi\rangle_{AC}\otimes|\Psi\rangle_{BC}$, where $|\Psi\rangle$ is any maximally entangled two-qubit state.  However, a strong monogamy relation of the form of Eq.~(\ref{chshmon}) holds if Alice is restricted to two fixed measurements~\cite{TV06}. We show here that the proof for this restricted case can be greatly simplified via  Eq.~(\ref{chshmon}). Moreover, the restriction can be removed entirely---both for CHSH and EPR-steering inequalities---when Alice's system is a qubit. 

The  simplified derivation proceeds as follows.  First, for fixed local measurements, with $\langle\mathcal B_{AB}\rangle$ denoting the expectation value of the CHSH operator corresponding to Alice and Bob's measurements on the shared  general tripartite  system, the set $R:=\{(\langle\mathcal B_{AB}\rangle,\langle\mathcal B_{AC}\rangle)\}$ over all states is convex, implying that its boundary is the envelope of a set of linear Bell inequalities of the form  $\langle \alpha\mathcal B_{AB}+\beta\mathcal B_{AC}\rangle\leq \gamma$.  Second,  saturation of  any such linear Bell inequality is achievable via projective measurements on a state having support on a single qubit for each party \cite{M05, TV06}.  Hence, from Eq.~(\ref{chshmon}) for the case $\bm a_1=\bm a'_1,\bm a_2=\bm a'_2$, we have
$R\subseteq \{ \langle\mathcal B_{AB}\rangle^2+\langle\mathcal B_{AC}\rangle^2\leq 8\}$ as desired. The tightness of the inequality is easily shown via an explicit example~\cite{TV06}.

Note that the second step of the above derivation corresponds to a significantly weakened form of Lemma~2 of Ref.~\cite{TV06}. In particular, it is not required that the observables and state are real relative to some basis set. Moreover, the nontrivial Lemmas~3 and~4 in Ref.~\cite{TV06} (which require analytically confirming, for arbitrary real pure three-qubit states, the equivalence of Eqs.~(5) and~(9) and the commutation relation following Eq.~(12) therein), are not needed at all.

Further, if Alice's system is a qubit, then the second step trivially goes through even if Alice chooses different pairs of directions for $\mathcal B_{AB}$ and $\mathcal B_{AC}$. It follows, via Eq.~(\ref{chshmon}), that at most one other party can violate a CHSH inequality with her qubit even if she optimises her measurement directions for each party.
This also allows us to obtain strong steering monogamy results for the steering of a qubit held by Alice, via local measurements by Bob and Charlie on arbitrary systems. For example, consider the powerful nonlinear steering inequality of Cavalcanti {\it et al.}~\cite{CJWR09} (with the roles of Alice and Bob reversed). As shown by Taddei {\it et al.}, any violation of this inequality implies Alice and Bob can also violate the CHSH inequality~\cite{TNA16}.  Hence, it is impossible for both Bob and Charlie to demonstrate steering of Alice's qubit via this inequality, even if she optimises her measurement directions in each case.

\paragraph{Robust statistical secret sharing.---}

Quantum secret sharing corresponds to coding classical or quantum information in an $n$-party state, such that it can only be recovered, via local measurements and classical communication, by $k>1$ of the parties~\cite{HBB99}.  In the ideal case only one measurement per party per state is required. In practice, however, unavoidable noise means the information will need to be recovered via the statistics of measurements on several copies of the shared state~\cite{HBB99}, corresponding what might be referred to as {\it statistical} secret sharing.  Moreover, alignment of distant reference frames (e.g., of polarisations) is in general required to  extract the information.

The anisotropic invariance property in Eq.~(\ref{aniso1}) suggests a variant of statistical secret sharing, in which any two parties can recover classical information encoded in a three-qubit state.  While this variant always requires several copies of the shared state, it is intrinsically robust to noise and invariant under rotations and reflections of local reference frames.  For example, suppose that the binary digit 0 is encoded in copies of a pure three-qubit state having eigenvalue gaps as per Eq.~(\ref{gaps}) with $g_1>g_2$ (i.e., $\delta s_2<0$). Similarly the digit 1 is encoded in copies of a second state with $g_1<g_2$ (i.e, $\delta s_2>0$). Any two parties can then recover this information by estimating $\mathcal S=TT^\top$ relative to arbitrary local basis sets, and thence the eigenvalue gaps $g_1$ and $g_2$. More information could be encoded in the values or ratios of $g_1$ and $g_2$, again with no reference frame alignment necessary, but would require more copies of each state to estimate the pairwise anisotropies to a sufficient accuracy. 

Further, if local isotropic noise (of different strengths) is applied to each qubit, then $g_1$ and $g_2$ are merely rescaled for each pair of qubits (see Supplemental Material~\cite{SM}). Hence, binary information encoded as above is preserved, i.e., it is robust to such noise. It would be of interest to compare  with the performance of other quantum secret sharing protocols in this regard, and for other noise models.

\paragraph{Quantum marginal problem.---}

Anisotropic invariance is also applicable to the quantum marginal problem, i.e., finding consistency constraints on sets of reduced density operators  \cite{K04}. For example, it is known that three qubit density operators $\rho_A$, $\rho_B$, $\rho_C$ are marginals of some pure three-qubit state if and only if their Bloch vector lengths satisfy $a+b\leq 1+c$ and its permutations~\cite{HSS03, K04}.  Our results imply that three two-qubit density operators, $\rho_{AB}$, $\rho_{AC}$, $\rho_{BC}$, can be marginals of such a state only if their spin anisotropies are identical and their  isotropic strengths  sum to 1.

\paragraph{Discussion.---}

The surprising invariance of pairwise anisotropy for three-qubit states leads to many interesting applications,  including a universal ordering of several quantum correlation measures, strong monogamy relations for various such measures,  tradeoffs between Bell nonlocality, 3-tangle and anisotropy, and a potentially useful statistical form of quantum secret sharing. 
It will be of significant further interest to extend the results and applications presented here, including to similar $k$-party invariants for $n$-qubit states and Gaussian states.

\paragraph{\bf Acknowledgements}~\\
 We thank  Lijun Liu, Li Li, Yu Xiang, Eric Cavalcanti and Howard Wiseman for helpful discussions. This research was supported by the ARC Centre of Excellence CE110001027.


%

\newpage

%

\onecolumngrid
\setcounter{page}{1}
\renewcommand{\thepage}{Supplemental Material -- \arabic{page}/5}
\setcounter{equation}{0}
\renewcommand{\theequation}{S\arabic{equation}}

\section{SUPPLEMENTAL MATERIAL}

\subsection{Isotropic vs anisotropic spin correlations}

For any given two-qubit state $\rho_{AB}$, the spin correlation matrix $T$---and the eigenvalues of the symmetric matrix $\mathcal{S}=TT^\top$ in particular---play a prominent role in assessing the suitability of the state for various quantum information tasks. 
Noting that $T\rightarrow O_A TO_B^\top$ under local orthogonal transformations of the Bloch sphere, i.e., rotations and reflections (corresponding to local unitary and antiunitary transformations of the qubits), it follows that $\mathcal{S}\rightarrow O_A	\mathcal{S}O_A^\top$, implying that the eigenvalues $s_1\geq s_2\geq s_3\geq0$ of $	\mathcal{S}$ are invariant under such transformations.  

As discussed in the main text, it turns out to be very useful to decompose $\mathcal S$ into its isotropic and anisotropic components:
\begin{align}
	\mathcal{S}=\mathcal{S}_{\rm iso}+\mathcal{S}_{\rm ani} 
	:=\frac{1}{3}\tr{\mathcal{S}}I_{3}+\left(\mathcal{S}- \frac{1}{3}\tr{\mathcal{S}}I_{3}\right),
\end{align}		
where $I_3$ denotes the $3\times3$ identity matrix. It follows that the isotropic strength in Eq.~(\ref{siso}) of the main text is given by
$s_{\rm iso} =\frac{1}{3}\tr{\mathcal{S}}=\frac{1}{3}\tr{\mathcal{S}_{\rm iso}}$, and that the differences $\delta s_j=s_j-s_{\rm iso}$ are given by the eigenvalues of $\mathcal{S}_{\rm ani}$.
Note that the invariant sum of the pairwise isotropic strengths in Eq.~(\ref{isotropy}) of the main text, for pure three-qubit states, corresponds to the matrix property
\beq
\mathcal{S}^{AB}_{\rm iso}+\mathcal{S}^{AC}_{\rm iso}+\mathcal{S}^{BC}_{\rm iso} = I_3 .
\eeq

\subsection{Proof of anisotropic invariance properties}

The main result of the paper, leading to the many applications therein, is the surprising fact that the spin anisotropy properties of each pair of qubits are identical for pure three-qubit states, i.e.,
\beq \label{aniso1SM}
\delta s^{AB}_j=\delta s^{AC}_j=\delta s^{BC}_j,\qquad j=1,2,3
\eeq
as per Eq.~(\ref{aniso1}) of the main text. This is equivalent to the three anisotropic matrices $\mathcal{S}^{AB}_{\rm ani}$, $\mathcal{S}^{AC}_{\rm ani}$, $\mathcal{S}^{BC}_{\rm ani}$ having identical eigenvalues, i.e., having the same characteristic polynomials:
\beq \label{polyn}
\det (\mathcal{S}^{AB}_{\rm ani}-\lambda I)= \det (\mathcal{S}^{AC}_{\rm ani}-\lambda I)=\det (\mathcal{S}^{BC}_{\rm ani}-\lambda I) .
\eeq
Now, if $\mathcal{S}_{\rm ani}$ denotes any one of these matrices, with eigenvalues $\delta s_j$, then the corresponding characteristic polynomial is given by
\begin{align}
P(\lambda) &= (\lambda-\delta s_1)(\lambda-\delta s_2)(\lambda-\delta s_3) \nn\\
&=\lambda^3  +(\delta s_1\delta s_2+\delta s_1\delta s_3+\delta s_2\delta s_3)\lambda - \delta s_1\delta s_2\delta s_3 \nn \\
&= \lambda^3  - \half(s_{\rm ani})^2\,\lambda - V_{\rm ani} .
\end{align}
Here the second line follows using $\sum_j\delta s_j=\tr{\mathcal{S}_{\rm ani}}=0$; the anisotropic strength and volume are defined by $s_{\rm ani}:= \big[\sum_j (\delta s_j)^2\big]^{1/2}$ and $V_{\rm ani}:=\delta s_1\delta s_2\delta s_3$ as per the main text; and the last line follows via
\begin{align} \nn
\delta s_1\delta s_2+\delta s_1\delta s_3+\delta s_2\delta s_3 = \half\big(\sum_j \delta s_j\big)^2-\half \sum_j (\delta s_j)^2 = -\half (s_{\rm ani})^2. 
\end{align}
Thus,  Eq.~(\ref{polyn}), and hence Eq.~(\ref{aniso1SM}), may be proved by showing that the anisotropic strengths and volumes are the same for each pair, for all pure three-qubit states $\ket{\psi_{ABC}}$. 

While the anisotropic measures $s_{\rm ani}$ and $V_{\rm ani}$ have simple definitions as per above in terms of the spin correlation matrix, they are extremely complicated when expressed in terms of the coefficients of $\ket{\psi_{ABC}}$ relative to some fixed basis. It is straightforward, nevertheless, to evaluate such expressions using a software package such as Mathematica, to find that indeed these measures are identical for each pair of qubits, immediately implying Eq.~(\ref{aniso1SM}) as desired (see also the next section for some explicit forms, including a more direct proof of the invariance of $s_{\rm ani}$).

\subsection{Explicit expressions for isotropic and anisotropic measures}

It is of interest to consider explicit expressions for isotropy and anisotropy measures for pure three-qubit states. This is done below for the isotropic and anisotropic strengths $s_{\rm iso}$ and $s_{\rm ani}$ for general states, as well as for the eigenvalues $s_j$ and volume $V_{\rm ani}$ in the case of W-class states \cite{DVC00}.
 
First, since the eigenvalues $s_j$, and hence the isotropic and anisotropic measures,  are invariant under local unitary operations, it is sufficient, and always possible, to transform a general pure state to the simplified form \cite{AACJLT00,AAER01}
\beq 
 \ket{\psi_{ABC}} = l_0\ket{000}+l_1e^{i\phi}\ket{100}+l_2\ket{101}+l_3\ket{110}+l_4\ket{111}, \label{cstate}
\eeq
with $l_j\geq 0$ and $\sum^4_{j=0}l^2_j=1$.  
This yields corresponding explicit expressions
\begin{align}
	\mf{a}&=(2l_0l_1\cos\phi, 2l_0l_1\sin\phi, 2l_0^2-1)^\top, \label{a} \\
	\mf{b}&=(2l_1l_3\cos\phi+2l_2l_4, -2l_1l_3\sin\phi,  1-2l^2_3-2l^2_4)^\top, \label{b} \\
	\mf{c}&=(2l_1l_2\cos\phi+2l_3l_4, -2l_1l_2\sin\phi,  1-2l^2_2-2l^2_4)^\top, \label{c}
\end{align}
for the  Bloch vectors, and 
\begin{align}
T^{AB}=	&\begin{pmatrix}
		2l_0l_3 & 0 & 2l_0l_1\cos\phi \\
		0 & -2l_0l_3 & 2l_0l_1\sin\phi \\
		-2l_1l_3\cos\phi-2l_2l_4 & 2l_1l_3\sin\phi & 1-2l_1^2-2l^2_2
	\end{pmatrix}, \label{ABtmatrix} \\ \nn
~\\
T^{AC}=	&\begin{pmatrix}
		2l_0l_2 & 0 & 2l_0l_1\cos\phi \\
		0 & -2l_0l_2 & 2l_0l_1\sin\phi \\
		-2l_1l_2\cos\phi-2l_3l_4 & 2l_1l_2\sin\phi & 1-2l_1^2-2l^2_3
	\end{pmatrix}, \label{ACtmatrix} \\ \nn
~\\
T^{BC}=	&\begin{pmatrix}
		2l_2l_3+2l_1l_4\cos\phi & -2l_1l_4\sin\phi & 2l_1l_3\cos\phi-2l_2l_4 \\
		-2l_1l_4\sin\phi & 2l_2l_3-2l_1l_4\cos\phi & -2l_1l_3\sin\phi \\
		2l_1l_2\cos\phi-2l_3l_4 & -2l_1l_2\sin\phi & 1-2l_2^2-2l^2_3
	\end{pmatrix} , \label{BCtmatrix} 
\end{align}
for the pairwise spin correlation matrices.

The corresponding isotropic strengths are found to have the relatively simple forms
\begin{align}
	s^{AB}_{\rm iso}&=\frac{1}{3}\left(1+8l^2_0l^2_3-4l^2_0l^2_2-4l^2_1l^2_4-4l^2_2l^2_3+8l_1l_2l_3l_4\cos\phi \right), \\
	s^{AC}_{\rm iso}&=\frac{1}{3}\left(1+8l^2_0l^2_2-4l^2_0l^2_3-4l^2_1l^2_4-4l^2_2l^2_3+8l_1l_2l_3l_4\cos\phi \right), \\
	s^{BC}_{\rm iso}&=\frac{1}{3}\left(1-4l^2_0l^2_2-4l^2_0l^2_3+8l^2_1l^2_4+8l^2_2l^2_3-16l_1l_2l_3l_4\cos\phi \right),
\end{align}
which clearly sum to 1 as per Eq.~(\ref{isotropy}) of the main text. 

Further, the anisotropic strength of the correlations between Alice and Bob is given by, using Eq.(\ref{ABtmatrix}),
\begin{align}
	& (s^{AB}_{\rm ani})^2
	 =\frac{2}{3}L_1^2+8L^2_2+8l^2_0l^2_1\left[l^2_0l^2_1+(-1+2l^2_1+2l^2_2+2l^2_3)\sin^2\phi\right],
\end{align}
 where
\begin{align*}
	L_1:=1-2l^2_0(3l^2_1+2l^2_2+2l^2_3)-4l^2_1l^2_4-4l^2_2l^2_3+8l_1l_2l_3l_4\cos\phi, ~~~
	L_2:=2l_0l_2l_3l_4+l_0l_1(-1+2l^2_1+2l^2_2+2l^2_3)\cos\phi.
\end{align*}
Since interchanging $B$ and $C$ corresponds to swapping $l_2$ and $l_3$ in Eq.~(\ref{cstate}) , it follows by inspection that $s^{AB}_{\rm ani}=s^{AC}_{\rm ani}$. By considering the alternative simplified form $\ket{\psi'_{ABC}} = l'_0\ket{000}+l'_1e^{i\phi'}\ket{001}+l'_2\ket{101}+l'_3\ket{011}+l'_4\ket{111}$,  obtained by swapping the roles of Alice and Charlie in Eq.(\ref{cstate}), one also has that $s^{CA}_{\rm ani}=s^{CB}_{\rm ani}$ (noting that $|\psi_{ABC}\rangle$ and $|\psi'_{ABC}\rangle$ are be related by local unitary transformations, under which the eigenvalues and hence the anistropic strengths are invariant). Hence, since the eigenvalues of $\mathcal{S}^{AB}=T^{AB}(T^{AB})^\top$ and $\mathcal{S}^{BA}=T^{BA}(T^{BA})^\top=(T^{AB})^\top T^{AB}$ are identical, it follows that $s^{AB}_{\rm ani}=s^{AC}_{\rm ani}=s^{BC}_{\rm ani}$  for general pure three-qubit states, as expected.
	
An explicit expression for the anisotropic volume, $V_{\rm ani}$, in terms of the wave function coefficients, may be calculated via any of  Eqs.~(\ref{ABtmatrix})-(\ref{BCtmatrix}), with the same result in each case. While the expression for the general case is too long and complicated to be usefully given here, anisotropy measures do have tractable forms for particular classes of states.

For example, a pure three-qubit state is a W-class state if and only if it is not biseparable and the 3-tangle vanishes~\cite{DVC00}.
Using the representative form for general states in Eq.~(\ref{cstate}), it is straightforward to calculate the concurrences of  each pair of qubits~\cite{W98}, yielding ${\mathcal C}^2(\rho_{AB})=4l_0^2l^2_3$, ${\mathcal C}^2(\rho_{AC})=4l^2_0l^2_2$, and ${\mathcal C}^2(\rho_{BC})=4l_2^2l_3^2+4l_1^2l^2_4-8l_1l_2s_3l_4\cos\phi$,  Consequently, the 3-tangle, $\tau$, is given by \cite{CKW00}
\begin{align}
	\tau&=1-a^2 - {\mathcal C}^2(\rho_{AB}) -  {\mathcal C}^2(\rho_{AC})
	=1-b^2 - {\mathcal C}^2(\rho_{AB}) -  {\mathcal C}^2(\rho_{BC}) 
	=1-c^2 - {\mathcal C}^2(\rho_{AC}) -  {\mathcal C}^2(\rho_{BC}) \label{perm} \\
	&=4l_0^2l^2_4. \label{tau}
\end{align}
 Thus, noting that $l_0=0$ yields a biseparable state in Eq.~(\ref{state}), it follows from Eq.~(\ref{tau}) that W-class states correspond to $l_4=0$ and $l_0>0$. 

The eigenvalues of the matrix $\mathcal{S}^{AB}=T^{AB}(T^{AB})^\top$ can be explicitly evaluated for W-class states, using Eq.~(\ref{ABtmatrix}) with $l_4=0$, with the result
\begin{align} \label{s1}
	s^{AB}_1&=\frac{1}{2}\left[(1-2l^2_2)^2+4(l^2_0l_3^2+l^2_1l^2_2) +\rt{(1-2l^2_2)^2+4(l^2_0l_3^2+l^2_1l^2_2)]^2-16l_0^2l^2_3(1-2l^2_2)^2}\right], \\ \label{s2}
	s^{AB}_2& =4l^2_0l^2_3, \\
	s^{AB}_3&=\frac{1}{2}\left[(1-2l^2_2)^2+4(l^2_0l_3^2+l^2_1l^2_2) -\rt{(1-2l^2_2)^2+4(l^2_0l_3^2+l^2_1l^2_2)]^2-16l_0^2l^2_3(1-2l^2_2)^2}\right].\label{s3}
\end{align} 
Further, the corresponding eigenvalues $s^{AC}_j$ and $s^{BC}_j$ are given by interchanging $l_2\leftrightarrow l_3$ and $l_2\leftrightarrow l_0$, respectively. It is straightforward to then calculate all functions of these eigenvalues explicitly.  For example, one obtains
\beq
V_{\rm ani} =\frac{2}{27}\left[ 1 -4(l_0^2l_2^2+l_0^2l_2^2+l_2^2l_3^2)\right] \left[ (1-2l^2_2)^2+4(l^2_0l_3^2+l^2_1l^2_2)]^2-16l_0^2l^2_3(1-2l^2_2)^2 +8l_0^2l_1^2l_2^2l_3^2\right]
\eeq
for the anisotropic volume of W-class states. 

One can similarly obtain relatively simple expressions for the class of states corresponding to $\phi=0$  in Eq.~(\ref{cstate}).

\subsection{ Independence of  3-tangle and anisotropy}

To demonstrate that the anisotropy properties of a state are independent of its 3-tangle, consider first the class of W-class states. As noted in the previous section, these states satisfy $\tau=0$.  However, it is clear from Eqs.~(\ref{s1})-(\ref{s3}) that the three eigenvalues $s^{AB}_j$ vary independently, and hence so do  the the two anisotropic measures $g_1=s^{AB}_1-s^{AB}_2$ and  $g_2=s^{AB}_2-s^{AB}_3$ in Eq.~(\ref{gaps}) of the main text.

Conversely, for the class of states defined by $l_0=\cos\alpha, l_1=l_2=l_3=\phi=0, l_4=\sin\alpha$ in Eq.~(\ref{cstate}), with $\alpha\in[0,\pi/2]$, one finds that $g_1$ and $g_2$ have fixed values independent of $\alpha$ (corresponding to $(s_{\rm ani})^2=2/3$ and $V_{\rm ani}=2/27$), whereas the 3-tangle follows from Eq.~(\ref{tau}) as $\tau=\sin^2 2\alpha$, and hence varies independently over all possible values.

\subsection{Entanglement ordering}

To obtain the quantitative entanglement-ordering relations for pure three-qubit states given in Eq.~(\ref{entangle ordering}) of the main text, note first that equating the purities of the reduced states $\rho_{AB}$ and $\rho_C$, and similarly for permutations of the qubits,
yields the expressions
\begin{align}
	s^{AB}_{\rm iso}=\frac{1}{3}\left(1+2c^2-a^2-b^2\right), \label{isoAB}\\
	s^{AC}_{\rm iso}=\frac{1}{3}\left(1+2b^2-a^2-c^2\right), \label{isoAC}\\
	s^{BC}_{\rm iso}=\frac{1}{3}\left(1+2a^2-b^2-c^2\right), \label{isoBC}
\end{align}
for the isotropic strengths.
These immediately lead to, for example, 
\beq
s^{AB}_{\rm iso}-s^{AC}_{\rm iso}=c^2-b^2.
\eeq
Hence, using anisotropic invariance as per Eq.~(\ref{aniso1}) of the main text, we obtain the crucial property
\beq
s^{AB}_j-s^{AC}_j=(\delta s^{AB}_j+s^{AB}_{\rm iso})-(\delta s^{AC}_j+s^{AC}_{\rm iso})=s^{AB}_{\rm iso}-s^{AC}_{\rm iso}=c^2-b^2 \label{crucial}
\eeq
for $j=1,2,3$.  Similar relations are obtained for each permutation of the parties A, B and C. 

It follows that  quantities involving the sum of eigenvalues of spin correlation matrices obey a similar relationship. For example, the Horodecki parameter ${\mathcal M}^{AB}=s^{AB}_1+s^{AB}_2$ and the fidelity of remote state preparation  ${\mathcal F}^{AB}=\half(s^{AB}_2+s^{AB}_3)$ satisfy
\beq
\half( {\mathcal M}^{AB}-{\mathcal M}^{AC})=\mathcal{F}^{AB}-\mathcal{F}^{AC}=c^2-b^2
\eeq
and its permutations.
Further, noting that the 3-tangle $\tau$ is invariant under permutations of the qubits, as per the equalities in Eq.~(\ref{perm}) \cite{CKW00}, comparison of each pair of these equalities yields
\begin{align}
	 {\mathcal C}^2(\rho_{AB})-{\mathcal C}^2(\rho_{AC})=c^2-b^2 
\end{align}
and its permutations.

Thus, we have derived the chains
\begin{align}
	b^2-c^2&=s_{\rm iso}^{AC}-s_{\rm iso}^{AC}={\mathcal C}^2(\rho_{AC})-{\mathcal C}^2(\rho_{AB})
	=\mathcal{F}^{AC}-\mathcal{F}^{AB}= \frac{1}{2}\left({\mathcal M}^{AC}-{\mathcal M}^{AB}\right), \label{b-c} \\
	a^2-b^2&=s_{\rm iso}^{BC}-s_{\rm iso}^{AC}={\mathcal C}^2(\rho_{BC})-{\mathcal C}^2(\rho_{AC})
	=\mathcal{F}^{BC}-\mathcal{F}^{AC}= \frac{1}{2}\left({\mathcal M}^{BC}-{\mathcal M}^{AC}\right), \label{a-b}\\
	c^2-a^2&=s_{\rm iso}^{AB}-s_{\rm iso}^{AC}={\mathcal C}^2(\rho_{AB})-{\mathcal C}^2(\rho_{BC})
	=\mathcal{F}^{AB}-\mathcal{F}^{BC}= \frac{1}{2}\left({\mathcal M}^{AB}-{\mathcal M}^{BC}\right), \label{c-a} 
\end{align}
of ordering relations, as per Eq.~(\ref{entangle ordering}) of the main text, as desired.  

Note that Eqs.~(\ref{b-c}) to~(\ref{c-a}) imply, in particular, that the triples $({\mathcal M}^{AB}, {\mathcal M}^{AC}, {\mathcal M}^{BC})$ and $(c, b, a)$ obey the same ordering, as per Eq.~(\ref{mab}) of the main text.  As a byproduct, this  generalises Theorems 1-4 of Ref.~\cite{PMC15}, and Theorem 2 of Ref.~\cite{SDSS16},  to all pure three-qubit states.

\subsection{Tradeoff relations}

To obtain the tradeoff relations for Bell nonlocality in Eqs.~(\ref{complementarity}) and~(\ref{trade}) of the main text, note first that for pure three-qubit states one has, using Eqs.~(\ref{nice}) and~(\ref{crucial}), 
\begin{align}
{\mathcal M}^{AB} = 1+s_3^{AB}-s_3^{AC}-s^{BC}_3 = 1 + (s_3^{AB}-s_3^{AC}) +(s_3^{AB}-s_3^{BC}) -s_3^{AB} 
=  1+2c^2-a^2-b^2-s^{AB}_3 . \label{intertrade}
\end{align}
Combining this expression with Eq.~(\ref{perm}) for the 3-tangle, and again using Eq.~(\ref{crucial}), then yields
\begin{align}
	{\mathcal M}^{AB}+\tau&= 1+2c^2-a^2-b^2-s^{AB}_3 
	+1-c^2-{\mathcal C}^2(\rho_{AC})-{\mathcal C}^2(\rho_{BC}) \nn \\
	&\leq  2+c^2-b^2-s^{AB}_3 \nn \\
	&=  2-s^{AC}_3 \leq  2 .
\end{align}
Similarly, permuting the parties gives ${\mathcal M}^{AC}+\tau\leq 2$ and ${\mathcal M}^{BC}+\tau \leq 2$. Thus, noting that the left hand sides of these inequalities are convex under mixing, the tradeoff relation~(\ref{complementarity}) follows for all pure and mixed 3-qubit states, confirming the conjecture in Ref.~\cite{PMC15}. The above derivation also confirms the validity of the conjecture made in Claim 2 of Ref.~\cite{PMC15} (noting that the geometric measure of entanglement equals $\max\{1-a^2,1-b^2,1-c^2\}$). In particular, from Eq.~(\ref{intertrade}) we have
\begin{align}
{\mathcal M}^{AB}+1-c^2
=2+c^2-a^2-b^2-s^{AB}_3 
=2-a^2+s^{AB}_3-s^{AC}_3-s^{AB}_3 = 2 -a^2 - s_3^{AC}
\leq 2,
\end{align}
where the second equality follows from Eq.~(\ref{crucial}) with $j=3$, and one similarly obtains the permutations
\begin{align}
	{\mathcal M}^{AC}+1-b^2=2 -c^2 - s_3^{BC}\leq 2,  \\
	{\mathcal M}^{BC}+1-a^2=2 -b^2 - s_3^{AB}\leq 2. 
\end{align}

Further, noting that the eigenvalue gaps in Eq.~(\ref{gaps}) satisfy $g_1+g_2=s^{AB}_1-s_3^{AB}$, it follows via Eq.~(\ref{nice}) of the main text that
\beq \label{tradeSM}
	{\mathcal M}^{AB} + g_1 + g_2 =1+s_1^{AB}-s_3^{AC}-s_3^{BC} \leq 1+s_1^{AB} \leq 2 ,
\eeq
for any pure 3-qubit state. Considering all permutations of the parties immediately yields the tradeoff relation in Eq.~(\ref{trade}) of the main text. A generalisation to mixed states is given in the next section.

\subsection{Recovering anisotropic invariance for mixed states via convex-roof extensions}

For a mixed three-qubit state, $\rho_{ABC}$, the matrices ${\mathcal S}^{AB}$, ${\mathcal S}^{AC}$, ${\mathcal S}^{BC}$ remain well defined, and hence the various measures of isotropy and anisotropy can be calculated for each pair of qubits. As noted in the main text, Eq.~(\ref{isotropy}) for the sum of the isotropic strengths generalises to the inequality
\beq
s^{\rm iso}_{AB}+s^{\rm iso}_{AC}+s^{\rm iso}_{BC}\leq 1
\eeq 
for such states.  However, aniostropic invariance as per Eq.~(\ref{aniso1}) is lost: the anisotropy properties will typically be different for each pair of qubits. 

This loss of anisotropic invariance for mixed states is not an issue for most of the relations in the main text, such as Eq.~(\ref{chshmon}), which have been derived for both pure and mixed states wherever indicated.  However, the tradeoff relation between anisotropy and Bell nonlocality in Eq.~(\ref{trade}) is an exception, being  derived  only for pure three-qubit states (see Eq.~(\ref{tradeSM}) in the previous section).  We show here that this relation can also be generalised to  mixed states via an appropriate convex-roof extension of  anisotropy measures.  More generally, such extensions also allow one to recover the property of anisotropic invariance for mixed states.

In particular, consider some measure of anisotropy, ${\cal Q}$, for two-qubit states, i.e, some fixed function of  $\delta s_1, \delta s_2, \delta s_3$. It follows from the anisotropic invariance property in Eq.~(\ref{aniso1}) of the main text that one has identical values,
\beq \label{same}
{\cal Q}^{AB}={\cal Q}^{AC}={\cal Q}^{BC} ,
\eeq
for any pure three-qbuit state $|\psi\rangle$. Denoting this common value by ${\cal Q}(|\psi\rangle\langle\psi|)$,  the
convex-roof extension of ${\cal Q}^{AB}$ to a mixed three-qubit state $\rho$ is given by
\beq
\tilde {\cal Q}^{AB}(\rho):=\min_{\{p_n, \ket{\psi_n}\}}\sum_n p_n Q^{AB}(\ket{\psi_n}\bra{\psi_n}) ,
\eeq
where the minimisation is over all mixtures $\rho=\sum_n p_n|\psi_n\rangle\langle \psi_n|$. Eq.~(\ref{same}) immediately generalises to 
\beq 
\tilde {\cal Q}^{AB}(\rho)=\tilde {\cal Q}^{AC}(\rho)=\tilde {\cal Q}^{BC}(\rho)  ,
\eeq
i.e., the convex-roof extension $\tilde {\cal Q}$ satisfies anisotropic invariance as desired.  

Although $\tilde {\cal Q}^{AB}(|\psi\rangle\langle\psi|)={\cal Q}^{AB}(|\psi\rangle\langle\psi|)$ for pure states, equality does not hold more generally.  However, by construction, one has the convexity property
\beq
\tilde {\cal Q}^{AB}(q\rho+q'\rho') \leq q\, \tilde {\cal Q}^{AB}(\rho) + q'\,\tilde {\cal Q}^{AB}(\rho')
\eeq
for any $\rho$, $\rho'$ and $q,q'\geq0$ satisfying $q+q'=1$.  This allows a simple generalisation of any given relation that is valid for the anisotropy of pure states.  For example, recalling that the Horodecki parameter ${\mathcal M}^{AB}$ is similarly convex \cite{HHH95}, one immediately obtains the generalisation
\beq 
\max \{{\mathcal M}^{AB}, {\mathcal M}^{AC}, {\mathcal M}^{BC}\}+\tilde g_1 + \tilde g_2 \leq 2  
\eeq
 of the tradeoff relation in Eq.~(\ref{trade}) of the main text,  valid for all mixed three-qubit states.

\subsection{Robustness of statistical secret sharing }

Finally, we demonstrate that the example of reference-frame-independent statistical secret sharing in the main text is robust to local isotropic noise.  In particular, such noise applied to a qubit state, $\rho_A$, is equivalent to mixing the state with the maximally-mixed state, corresponding to a completely positive map of the form
\beq
\phi_A(\rho_A) := \eta_A \,\rho_A + (1-\eta_A)\,\half \mathbbm{1}_A
\eeq
for some $0\leq \eta_A \leq 1$.  Hence, if $T^{AB}$ is the spin correlation matrix for the three-qubit state $\rho_{AB}$, then the spin correlation matrix corresponding to $(\phi_A\otimes\phi_B)(\rho_{AB})$ is easily found to be given by $\eta_A\eta_BT^{AB}$.  It immediately follows that ${\cal S}^{AB}$ scales as $(\eta_A\eta_B)^2$ under such noise, and hence its eigenvalues scale in the same way. Thus, the eigenvalue gaps $g^{AB}_1$ and $g^{AB}_2$ in Eq.~(\ref{gaps}) of the main text are similarly rescaled, implying that the sign of their difference is invariant under such noise, as claimed. Note that the ratios of the gaps are similarly invariant.


\end{document}